\documentclass[prr,twocolumn,amsmath,amssymb,amsthm,superscriptaddress]{revtex4-2}

\usepackage{graphicx,amsmath,relsize,epstopdf,upgreek,color,mathtools,bm,times,braket,rotating,tabularx,booktabs,setspace}
\usepackage[colorlinks=true,linkcolor=blue,citecolor=blue,urlcolor =blue]{hyperref}
\usepackage[hyphenbreaks]{breakurl}
\usepackage[T1]{fontenc}

\newcommand{\CORR}[1]{\textcolor{red}{#1}}

\newcommand{\inner}[2]{\langle #1|#2\rangle}
\newcommand{\opinner}[3]{\langle #1|#2|#3\rangle}

\newcommand{\tr}[1]{\mathrm{tr}\left\{#1\right\}}

\newcommand{\I}{\mathrm{i}}

\newcommand{\E}[1]{\mathrm{e}^{\mbox{\footnotesize$#1$}}}

\newcommand{\sech}{\mathrm{sech}}

\newcommand{\ML}{\widehat{\varrho}_\textsc{ml}}
\newcommand{\PR}{\mathrm{pr}}
\newcommand{\RB}{\mathrm{RB}}
\newcommand{\deff}{d_{\mathrm{eff}}}
\newcommand{\dRB}{d_{\mathrm{RB}}}

\newcommand{\HERM}[2]{\mathrm{H}_{\,#1}\!\left(#2\right)}

\newcommand{\appropto}{\mathrel{\vcenter{
			\offinterlineskip\halign{\hfil$##$\cr
				\propto\cr\noalign{\kern2pt}\sim\cr\noalign{\kern-2pt}}}}}

\begin{document}

\title{Evidence-based certification of quantum dimensions}

\author{Y. S. Teo}
\email{ys\_teo@snu.ac.kr}
\affiliation{Department of Physics and Astronomy, 
	Seoul National University, 08826 Seoul, South Korea}
	
\author{S. U. Shringarpure}
\email{saurabh.s@snu.ac.kr}
\affiliation{Department of Physics and Astronomy, 
	Seoul National University, 08826 Seoul, South Korea}

\author{H. Jeong}
\email{h.jeong37@gmail.com}
\affiliation{Department of Physics and Astronomy, 
	Seoul National University, 08826 Seoul, South Korea}
	
\author{N. Prasannan}
\affiliation{Integrated Quantum Optics Group, Applied Physics, University of Paderborn, 33098 Paderborn, Germany}

\author{B. Brecht}
\affiliation{Integrated Quantum Optics Group, Applied Physics, University of Paderborn, 33098 Paderborn, Germany}

\author{C. Silberhorn}
\affiliation{Integrated Quantum Optics Group, Applied Physics, University of Paderborn, 33098 Paderborn, Germany}

\author{M. Evans}
\email{mevansthree.evans@utoronto.ca}
\affiliation{Department of Statistical Sciences, University of Toronto, Toronto, Ontario, M5S 3G3, Canada}

\author{D. Mogilevtsev}
\affiliation{B. I. Stepanov Institute of Physics, NAS of Belarus, Nezavisimosti ave. 68, 220072 Minsk, Belarus}

\author{L. L. S{\'a}nchez-Soto}
\affiliation{Departamento de {\'O}ptica, Facultad de F{\'i}sica, Universidad Complutense, 28040 Madrid, Spain}
\affiliation{Max-Planck-Institut f{\"u}r die Physik des Lichts, Staudtstra{\ss}e 2, 91058 Erlangen, Germany}

\begin{abstract}
	Identifying a reasonably small Hilbert space that completely describes an unknown quantum state is crucial for efficient quantum information processing. We introduce a general dimension-certification protocol for both discrete and continuous variables that is fully evidence-based, relying solely on the experimental data collected and no other unjustified assumptions whatsoever. Using the Bayesian concept of relative belief, we take the effective dimension of the state as the smallest one such that the posterior probability is larger than the prior, as dictated by the data. The posterior probabilities associated with the relative-belief ratios measure the strength of the evidence provide by these ratios so that we can assess whether there is weak or strong evidence in favor or against a particular dimension. Using experimental data from spectral-temporal and polarimetry measurements, we demonstrate how to correctly assign  Bayesian plausible error bars for the obtained effective dimensions.  This makes relative belief a conservative and easy-to-use model-selection method for any experiment.
\end{abstract}

\maketitle

\emph{Introduction.}---The  stunning progress of modern quantum technologies ultimately relies on the ability to create, manipulate, and measure quantum states. All of these tasks require a careful verification of their quality. 

Given experimental data, a first crucial step is to estimate the dimension of the physical system (which, loosely speaking, is the number of significant degrees of freedom) without making spurious unvalidated assumptions about the devices used in the experiment. Such kind of dimensionality assessment, for instance, is necessary for executing quantum technologies~\cite{Plastino:2015aa} and has, consequently, received a lot of attention from various perspectives~\cite{Brunner:2008aa,Wehner:2008aa,Gallego:2010aa,Hendrych:2012aa,Ahrens:2012aa,DallArno:2012aa,Brunner:2013aa,Guhne:2014aa,Bowles:2014aa,DAmbrosio:2014aa,Ahrens:2014aa,Sun:2016aa,Cai:2016aa,Cong:2017aa,Sun:2020aa,Sohbi:2021aa}. 

Estimating the dimension of an unknown quantum system has also been important for a number of applications in quantum information~\cite{Lanyon:2009aa,Spekkens:2001aa,Massar:2002aa,Molina-Terriza:2005aa,Duclos-Cianci:2013aa,Campbell:2014aa,Luo:2019aa,Duclos-Cianci:2013aa}. As an example, proper dimension analysis permits security proofs of certain cryptographic schemes~\cite{Pawowski:2011aa,Woodhead:2015aa,Woodhead:2016aa,Goh:2016aa,Primaatmaja:2023aa} and has also been used for randomness certification~\cite{Lunghi:2015aa,Miao:2022aa}. Hilbert-space model-dimension certification is especially relevant to quantum tomography, where one assumes a fixed dimension, and it is therefore clear that the choice of a small Hilbert space that fully contains the quantum state~$\varrho$ can significantly reduce computational overheads. 

For discrete-variable~(DV) systems, if~$\varrho$ is sparse in some basis, then finding a small Hilbert space that contains~$\varrho$ can combat the exponentially-growing processing and storage complexities with the qubit number. For continuous-variable~(CV) systems, where a physical~$\varrho$ naturally has vanishing photon-number distribution tails, ascertaining the correct supportive Hilbert space would evidently avoid numerical artifacts originating from an erroneous truncation. Methods for Hilbert-space truncation~\cite{Mogilevtsev:2017aa,Teo:2016aa} employing commuting measurements and the maximum-likelihood principle~\cite{Banaszek:1999ml,Fiurasek:2001mq,lnp:2004uq,Rehacek:2007ml,Teo:2011me,Teo:2015qs} work well in the regime of very large datasets, but do not flexibly permit the assessment of additional prior knowledge about~$\varrho$ one might already have.

For a low-rank or nearly-pure~$\varrho$, using fewer measurement outcomes that are still \emph{informationally complete}~(IC) for state reconstruction is possible with modern compressive tomography~\cite{Gil-Lopez:2021aa,Teo:2021aa,Teo:2021ab,Teo:2020cs,Kim:2020aa,Gianani:2020aa,Ahn:2019aa,Ahn:2019ns} that does not rely on any assumptions or measurement restrictions (but still requires the system dimension), a leap from traditional compressed-sensing schemes~\cite{Gross:2010cs,Kalev:2015aa,Goyeneche:2015aa,Baldwin:2016cs,Steffens:2017cs}.

In this work, we propose a Bayesian evidence-based dimension-certification scheme that works for \emph{all} quantum systems, measurements and datasets of any sample size with no statistically-unjustified assumptions or restrictions whatsoever. It is built on the data-driven and irrefutable logic of the \emph{relative belief}~(RB) reasoning~\cite{Evans:2015aa,Evans:2016aa,Evans:2018aa,Al-Labadi:2018aa,Nott:2020aa,Nott:2021aa,Englert:2021aa,Evans:2023aa} to unambiguously assert the \emph{plausible} Hilbert spaces containing, or, more generally, the plausible hypotheses about~$\varrho$ that are supported by the data---the posterior probability must exceed the prior probability. Its first utility in quantum information gave rise to novel types of error-region constructions~\cite{Shang:2013cc,Li:2016da,Teo:2018aa,Oh:2018aa,Oh:2019aa,Oh:2019ab,Sim:2019aa} and statistically unambiguous refutation of local hidden-variable models for qubit systems~\cite{Gu:2019very}.

We shall now present a \emph{relative-belief dimension certification}~(RBDC) scheme that uniquely determines the smallest Hilbert space containing~$\varrho$ according to what the data tell us. In this Bayesian framework, all dimensions certified by RBDC are naturally endowed with error bars corresponding to given credibilities. With real experimental data, we demonstrate that RBDC can be routinely carried out with no technical difficulties. 

\emph{Relative-belief dimension certification.}---For a given positive operator-valued measure~(POVM)---the operator set $\{\Pi_j\geq0\}$ such that $\sum_j\Pi_j=1$---we would like to ascertain if an unknown state~$\varrho$ can be fully contained in a Hilbert space of an effective dimension~$\deff$. As~$\varrho$ is unknown and to be estimated from data, this translates to the operational question ``What is~$\deff$ for the maximum-likelihood~(ML) estimator~$\ML$ given the POVM data $\mathbb{D}$?'' The first step in RBDC would be to decide the largest dimension~$D$ one wishes to examine and assign a set of~$D-1$ \emph{prior probabilities}~$\PR(d)$ for~$2\leq d\leq D$. These represent our initial takes on how probable it is that a Hilbert space of dimension~$d$ contains~$\varrho$, which are regarded as \emph{intuitive assumptions to be tested}. 

A proper Hilbert-space truncation entails a good choice of a sparse basis representation for $\varrho$. In the context of dimension certification for multiphoton sources of limited intensity, for instance, an appropriate choice would be the standard computational basis~(or the Fock basis). After the experiment, IC data~$\mathbb{D}$ collected permits us to calculate the \emph{posterior probability} $\PR(d|\mathbb{D})=L_d\,\PR(d)/\sum^D_{d^{\prime}=1}L_{d^{\prime}}\PR(d^{\prime})$ for each~$d$, where $L_d\equiv L(\mathbb{D}|\ML(d))$ is the likelihood of obtaining~$\mathbb{D}$ from the $d$-dimensional~$\ML(d)$. The RB criterion~\cite{Evans:2015aa} states that a value of~$d$ is \emph{plausible} based on evidence from~$\mathbb{D}$ when the RB~ratio
\begin{equation}
	\RB(d)=\frac{\PR(d|\mathbb{D})}{\PR(d)}=\frac{L_d}{\sum^D_{d^{\prime}=1}L_{d^{\prime}}\, \PR(d^{\prime})}>1\,.
	\label{eq:RBcrit}
\end{equation} 
Physically, a~$\RB(d)>1$ reflects that $\mathbb{D}$ indeed supports the supposition that a $\dRB$-dimensional Hilbert space supports the unknown~$\varrho$, as a larger posterior probability relative to the prior probability strengthens our initial belief after-the-fact. As a side note, the concept of RB is naturally compatible with that of ML, as $\ML(d)$ may also be regarded as the maximum-RB estimator for a given $d$. 

Criterion~\eqref{eq:RBcrit} presents an unambiguous measure of statistical evidence, as the unit cut-off imposed by the posterior-prior ratio originates from the inherent definitions of these probabilities as belief measures themselves with a solid statistical foundation. This is unlike the $p$-value, for instance, that not only lacks the ability to find evidence in favor of any hypothesis~(it can only make a case \emph{against} some hypothesis), but is also, in general, an invalid measure of evidence in statistics~\cite{Benjamin:2018redefine,Wasserstein:2016ASA,Evans:2015aa}. The RB criterion~\eqref{eq:RBcrit} also permits a more direct evaluation of data evidence in favor or against any hypothesis without additional arbitrary specification of error probabilities for wrongfully accepting or refuting a hypothesis in regular hypothesis-testing paradigms. Rather, this technical aspect is, expectedly, built into the RB machinery, which is discussed in the companion technical article~\cite{Teo:2024relative-belief}.

\begin{figure}[t]
	\includegraphics[width=\columnwidth]{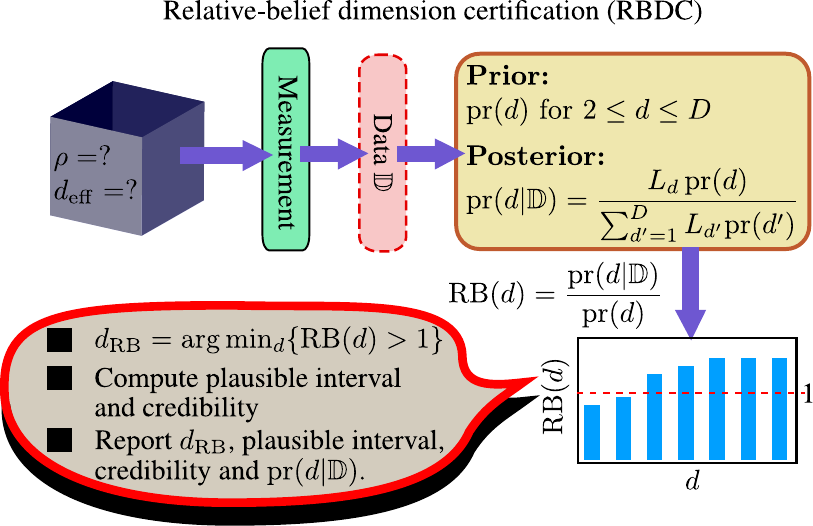}
	\caption{\label{fig:scheme}Schematic RBDC for general quantum tomography.}
\end{figure}

The procedure of RBDC, summarized in Fig.~\ref{fig:scheme}, is to take the smallest~$d\equiv\dRB$ that satisfies condition~\eqref{eq:RBcrit} as~$\deff$. It is clear that neither additional assumption about~$\varrho$ nor any other quantity is needed to carry out dimension certification with RB---everything is encoded in the dataset~$\mathbb{D}$ waiting to be extracted. It is straightforward to see that for a~$\mathbb{D}$ of very large sample-copy number~$N$, the influence of each likelihood function~$L_d$ overwhelms~$\PR(d)$, so that in this data-dominant situation, $\RB(d)$ is essentially governed by~$L_d$. As a larger Hilbert space surely contains the ML estimator derived from a smaller space, we have the monotonicity property~$L_{d}\leq L_{d+1}$, and hence \mbox{$\RB(d\geq\dRB)>1$} in the asymptotic limit. Evidently, the pathological prior that preferentially picks some~$d=d_0$, that is the Kronecker-delta prior~$\PR(d)=\delta_{d,d_0}$, ignores all data and must, of course, be excluded in any scientific inquiry. Under such a statistically natural and logical Bayesian system, where initial beliefs may be accepted or refuted solely by~$\mathbb{D}$ without the need for \emph{ad hoc} criteria, any quantum tomography procedure may be carried out through RBDC without spurious assumptions, including the insistence of~$\deff$, using the same tomographic datasets that reconstruct~$\varrho$.

Moreover, the posterior probability quantifies the strength of our belief that~$\deff=\dRB$ whenever $\RB(\dRB)>1$. If we think that this magnitude is too small, then we are free to choose another~$\deff$ that gives a larger RB~ratio, which is a feature that is fundamentally lacking in usual hypothesis testing~\cite{Evans:2015aa}. 

The tools for assessing the credibility of~$\dRB$ are built into the Bayesian character of RBDC. In particular, we can assign a \emph{plausible interval}~$[\dRB,\dRB+\Delta]$ for some integer~$\Delta$ of credibility~$\mathcal{C}_\Delta=\sum^{\dRB+\Delta}_{d'=\dRB}\PR(d'|\mathbb{D})\leq1$, which is the conditional probability~(given $\mathbb{D}$) that this interval contains all plausible~$d$ such that $\RB(d)>1$. Since $L_d$ is monotonic in~$d$, it is clear that all $d\geq\dRB$ will be plausible for a large~$N$. In this way, $\dRB$ comes with natural error intervals endowed by the Bayesian RB~framework. Operation-wise, $|L_d|$ can still be very small even for moderate values of~$N$. Nonetheless, routine computation of all these elements is now possible with recent developments in storing ultra-high precision formats, ideal for coping with minuscule $L_d$~values~\cite{DErrico:2018hpf}.  Thus, the result of RBDC is the report of the ML estimator, $\dRB$ and its plausible interval, the corresponding~\CORR{$\mathcal{C}_\Delta$}s and posterior probabilities.

Owing to its versatility, the RB paradigm applies to the assessment of very general hypotheses concerning a given quantum system. For the purpose of RBDC, we note that practically any data corresponding to a POVM that measures $\varrho$ are applicable. This POVM need not even be IC, that is, a unique reconstruction of the full $\varrho$ is not necessary; the simplest set of von~Neumann orthonormal projectors that probe diagonal elements of~$\varrho$ (the source's photon-number distribution, for instance) suffice for RBDC, as the sparsity of diagonal elements define the encapsulating Hilbert space of $\varrho$ as a consequence of the positivity~constraint.

\begin{figure}[t]
	\includegraphics[width=\columnwidth]{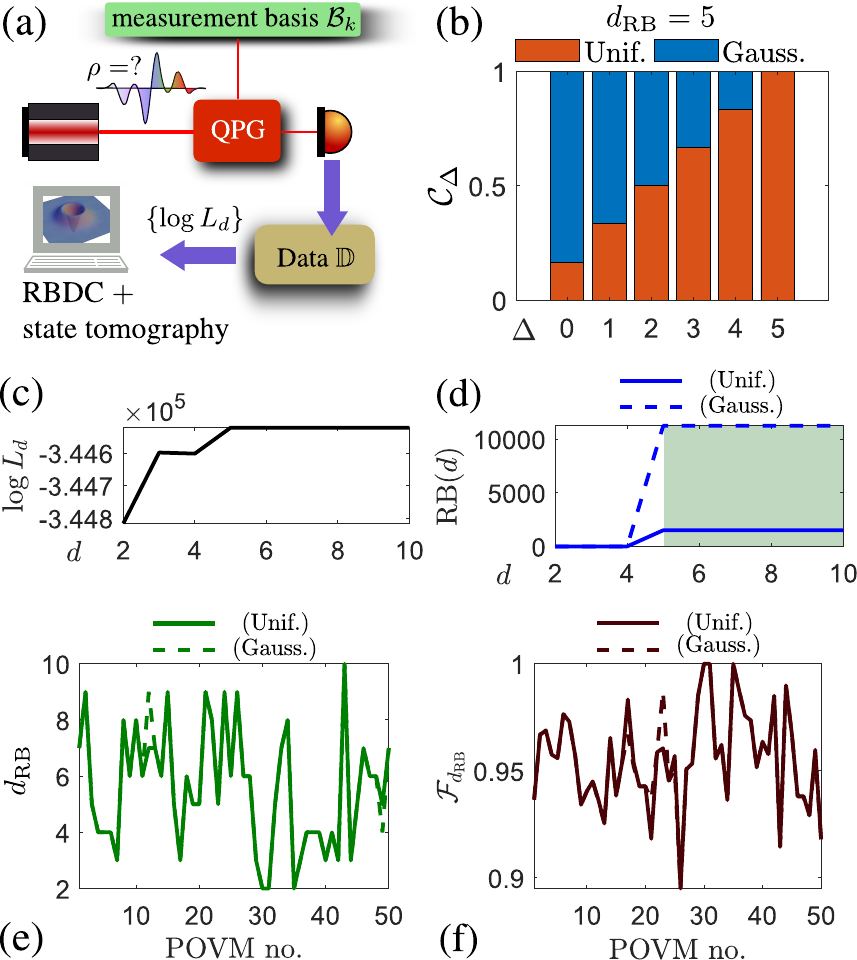}
	\caption{\label{fig:TF_res}v(a)~The experimental scheme for RBDC tomography on a degree-one temporal Hermite--Gaussian~(HG$_1$) quantum state consists of measuring $K=11$ von~Neumann bases~$\{\mathcal{B}_1,\mathcal{B}_2\ldots \mathcal{B}_K\}$ under random unitary rotations realized by the~QPG, which form one IC~POVM. The priors considered for RBDC are uniform~[$\PR(d)=1/(D-1)$] and Gaussian~[$\PR(d)\propto\exp(-(d-2)^2)$], the latter of which reflects a stronger conviction that $\dRB=2$ is sufficient to describe~$\varrho$. We take~\mbox{$D=10$}.
	(b)~The credibilities~$\mathcal{C}_\Delta$ for the plausible intervals $[\dRB,\dRB+\Delta]$ with $0\leq\Delta\leq2$ are plotted for one such POVM ($N\sim10^4$ and $\dRB=5$), where its corresponding (c)~log-likelihood and (d)~log-RB ratio follow the same trend over~$d$ as $N$ is rather large [plausible interval of~(b) shaded]. (e,f)~Data evidence from 50 randomly generated POVMs shows highly similar RBDC results and state-reconstruction fidelities~$\mathcal{F}_{\dRB}$ with both prior distributions for these large datasets.}
\end{figure}

\emph{RBDC in spectral-temporal state tomography.}---A crucial application of RBDC is quantum tomography of all physical~CV systems residing in an infinite-dimensional Hilbert space, but possessing a finite photon-number support. One promising platform that utilizes CV tomography is time-frequency encoding~\cite{Brecht:2015aa,Sergei:2019qip,Gil-Lopez:2021aa}, which serves as an alternative for a plethora of quantum-information and communication tasks~\cite{Zhou:2003aa,Eckstein:2011aa,Babazadeh:2017aa,Cozzolino:2019aa,Raymer:2020aa}.

In the experiment, we generate states whose information is encoded on Hermite--Gaussian~(HG) temporal modes, that is, field-orthogonal pulses with HG-shaped complex spectral amplitudes. We then execute a randomized compressive tomography on these states using a quantum pulse gate (QPG). The QPG is a dispersion-engineered sum-frequency generator in a periodically poled lithium niobate waveguide~\cite{Brecht:2014aa}. It combines a single-photon-level signal at telecom wavelengths and a strong classical pump pulse at around 860~nm. By shaping the complex amplitude of the pump pulse, the user defines the temporal mode that is selected by the QPG. The part of the signal that exhibits field overlap with this mode is up-converted and the converted output is detected with a single photon counter. A successful detection event implements a POVM onto the temporal mode defined by the pump~\cite{Ansari:2017aa}. Collecting counts for a set amount of time and for a set of IC pump temporal modes then allows for assessing the effective dimension of the state via RBDC.

Figure~\ref{fig:TF_res} showcases the positive performance of RBDC in determining the correct $\dRB$ of a temporal HG state $\varrho=\ket{\mathrm{HG}_n}\bra{\mathrm{HG}_n}$~($n=1$ for our experiments), with the temporal wave~function~$\inner{t}{\mathrm{HG}_n}\propto\E{-t^2/2}\HERM{n}{t}$. All datasets obtained from the QPG possess persistent background noise caused by detector dark counts, leading to a very small bias towards diagonal  matrix elements in $\ML(d)$ that do not vanish for any~$d$. We remove this bias by directly incorporating the constraint that all nonzero diagonal elements of $\ML(d)$ are $>0.01$ \emph{directly} into the~ML reconstruction, in addition to the positivity constraint, by following modified projected gradients~\cite{Shang:2017sf}. Such a noise subtraction may be viewed as additional prior information that enters the posterior computation for RBDC. In the~SM, we present the original data with background noise, where all $\deff$s are typically large.

\begin{figure}[t]
	\includegraphics[width=\columnwidth]{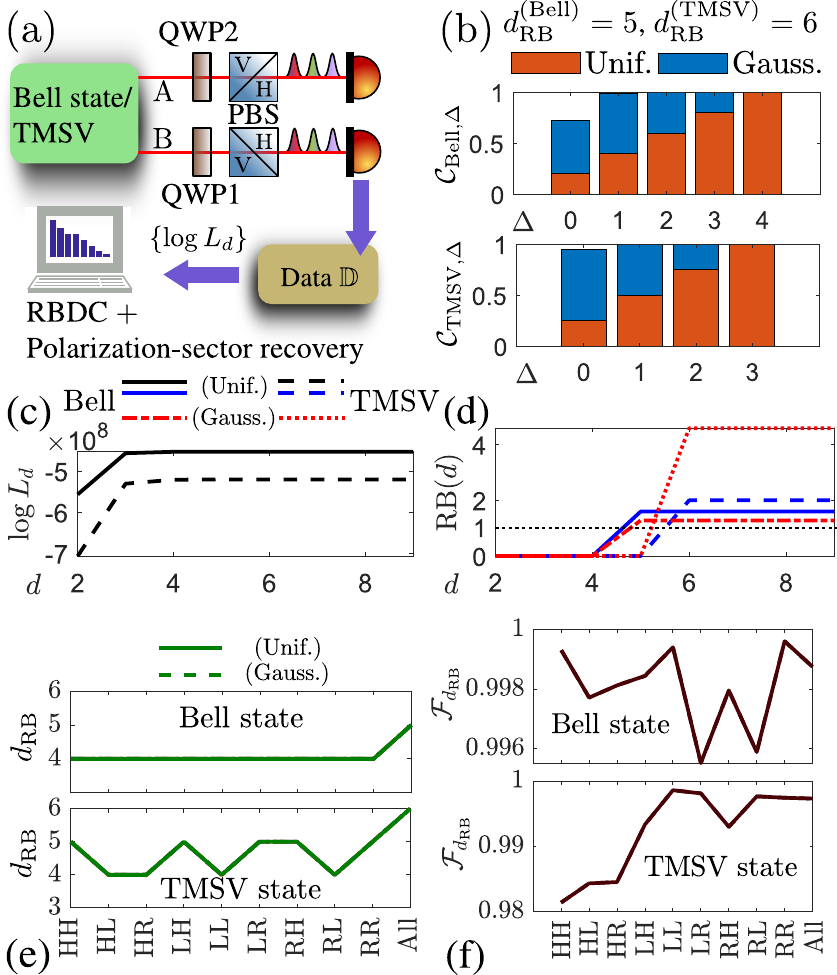}
	\caption{\label{fig:QPol_res}(a)~The quantum polarimetry setup involves a source of either Bell or TMSV states respectively generated at $r=$1.96~dB and 2.12~dB. The QWP angles are set to correspond the two-arm polarization settings \textsc{hh}, \textsc{hl}, \textsc{hr}, \textsc{lh}, \textsc{ll}, \textsc{lr}, \textsc{rh}, \textsc{rl} and \textsc{rr}, where \textsc{h}, \textsc{l} and \textsc{r} are the standard horizontal, left- and right-circular polarizations. Here, $\dRB$ refers to the effective mode dimension in \emph{each} arm. We have $D=9$ and the Gaussian prior~$\propto\exp[-(d-5)^2]$. (b)~Plots of~$\mathcal{C}_\Delta$ for Bell- and TMSV-state plausible intervals for the combined datasets of all nine measurement settings are shown. (c,d)~This time, $N\sim 10^8$ per setting, so that the overwhelming~$\mathbb{D}$ has very low data noise and $L_d$ increases monotonically with~$d$. (e,f)~We denote by $\mathcal{F}_{\dRB}=\sum^{\dRB-1}_{m,n=0}\sqrt{p_{mn}\widehat{p}_{\mathrm{ML},mn}}$  the Bhattacharyya fidelity between diagonal elements (or photon-number distributions $p_{mn}$) of the true reduced state~$\varrho'$ and ML estimator~($\widehat{p}_{\mathrm{ML},mn}$) \emph{extended} to the $D$-dimensional Hilbert space. The RBDC performances with both priors are identical in such data-dominant situations.}
\end{figure}

\emph{RBDC in quantum polarimetry.}---Apart from full quantum-state tomography, RBDC is also applicable for estimating key properties of a quantum source. One such scenario is quantum polarimetry for high-dimensional quantum systems to investigate polarization sectors~\cite{Goldberg:2021aa}, which encode interesting hidden quantum properties beyond the classical polarization description~\cite{Klyshko:1992aa,Klyshko:1997aa,Tsegaye:2000aa,Usachev:2001aa,Luis:2002aa,Agarwal:2003aa,Sanchez-Soto:2007aa,Bjork:2010aa,Sanchez-Soto:2013aa,Bjork:2015aa,Bjork:2015ab,Shabbir:2016aa,Goldberg:2017aa,Bouchard:2017aa,Goldberg:2020aa} and shall reveal the versatility of RBDC in certifying dimensions of~DV systems. A typical polarimetry setup consists of wave plates and beam splitters that transform an incoming optical signal of unknown state~$\varrho$. Coupled with photon-number-resolving detectors~(PNRDs) of $\approx 90\%$ efficiency, these passive components realize POVMs that only probe the polarization sector of~$\varrho$ that is smaller than the complete $\dRB^2$-dimensional state space. Nevertheless, quantum polarimetry is an economical method to study the polarization sectors, which include photon-number statistics.

The polarimetry setup is operated by a dispersion-engineered periodically-poled potassium titanyl phosphate (PPKTP) waveguide  inside a Sagnac interferometer~\cite{Meyer-Scott:2018nonlinear,Prasannan:2022aa}. Type-II parametric down-conversion with significant multiphoton contributions is generated in the clockwise and counter-clockwise direction. A folded~$4f$ spectrometer is used to select a narrow part of the pump-laser spectrum (100~fs pulse duration and 774~nm central wavelength), yielding pump pulses of 1~ps duration. Interfering the resulting spectrally-decorrelated signal and idler beams on the Sagnac polarizing beam splitter~(PBS) generates either a Bell state [$\ket{\text{Bell}}\propto\sum^\infty_{m,n=0}\ket{m,n-m,n-m,m}(\tanh r)^{n}$] or a two-mode squeezed vacuum~(TMSV) state [$\ket{\text{TMSV}}\propto\sum^\infty_{m,n=0}\ket{n-m,n-m,n,n}(\tanh r)^{n}$], where $r$~is the squeezing parameter, $\ket{m,n,m',n'}=\ket{m'}_{\text{A},\textsc{h}}\ket{n'}_{\text{A},\textsc{v}}\ket{m'}_{\text{B},\textsc{h}}\ket{n'}_{\text{B},\textsc{v}}$. The quarter-wave plates~(QWP) on arms~A and B \emph{each} realizes $U_\varphi= \exp[- \I \varphi(a^\dag_\textsc{h}a_\textsc{h}-a^\dag_\textsc{v}a_\textsc{v})/2]$. The PNRDs each has a maximum photon-number resolution of $n_0=8$, so that the maximum testable dimension is $D=n_0+1=9$. These PNRDs, therefore, function as $n_0=8$ ``on-off'' detectors~\cite{Sperling:2012true}. As $U_\varphi$ is an element of the SU(2) group under the Jordan--Schwinger representation~\cite{Jordan:1935Zusammenhang,Schwinger:1952report}, the complete polarimetry POVM for each spatial mode probes \emph{only} the polarization sector and not the complete state space~\cite{Goldberg:2021aa}. Consequently, measuring only the \textsc{h}~port on each arm (while tracing out the other \textsc{v}~port) results in an overall POVM with only diagonal outcomes in the computational basis. This implies that the estimable state elements are only the diagonal ones $p_{mn}\equiv\opinner{m,n}{\varrho'}{m,n}=(\sech\,r)^4\,(\tanh r)^{2(m+n)}$ of $\varrho'=\mathrm{tr}_{\text{A},\textsc{v},\text{B},\textsc{v}}\{\ket{\text{Bell or TMSV}}\bra{\text{Bell or TMSV}}\}$.

Figure~\ref{fig:QPol_res} shows the feasibility of carrying out RBDC for data~$\mathbb{D}$ of very large copy numbers that are free of background noise. In such data-dominant scenarios, the prior choice becomes irrelevant and RBDC is heavily influenced by~$\mathbb{D}$. For such huge datasets that are also almost void of statistical or systematic errors, the fact that $L_d$ is so strongly peaked means that, almost always, $\mathrm{RB}(d)>1$ only when $L_{\dRB}$ is the true maximum value $L_{d_\mathrm{c}}$ achieved by a~$d$ greater than or equal to some critical dimension~$d_\mathrm{c}\leq D$. 

The results are a testament to the possibility of computing posterior probabilities from such extremely small values of~$L_d$ without running into numerical underflow problems. We owe this to modern ways of storing extremely tiny numbers, which permits routine usage of this highly reliable and logical Bayesian machinery for dimension certification on very general and realistic experimental data. All numerical values of~$L_d$ and posterior probabilities used in Figs.~\ref{fig:TF_res}(b) and~\ref{fig:QPol_res}(b) are tabulated in~SM, with storage precision as high as $10^{-306872481}$.

\emph{Discussion.}---We presented a completely evidence-based dimension-certification protocol that only extracts information from the experimental dataset obtained to ascertain the smallest effective dimension that fully contains the given unknown quantum state for a chosen truncation basis. This scheme is based on the powerful method of relative belief that quantitatively assesses how well the dataset supports our belief that the state should be contained by a Hilbert space by comparing the posterior probability of this belief with the initial prior probability before the experiment was conducted.

We tested this protocol with real experimental data obtained from spectral-temporal and polarimetry measurements and demonstrated that relative-belief ratios and credibility computations are now a reality using modern numerical toolboxes that competently handle extremely small numbers without underflow issues. By trusting only the data and nothing else, our protocol never oversteps its boundaries and conclude a Hilbert space of an overly optimistic (smaller) dimension that is unjustified by the data. Thus, our work sets a concrete example of what a truly evidence-based prescription that is simple and feasible for any quantum experiment should be.

Note that relative-belief dimension certification, consequently, also does not assume a specific truncation basis used. If there is strong conviction that the unknown state is sparse under a privileged basis choice, then this basis may be used with this scheme to further reduce the obtained effective dimension. On the other hand, the protocol readily informs us when this conviction is wrong by yielding posterior probabilities smaller than the corresponding prior ones. Such a basis choice is but one of many types of prior knowledge one can incorporate in the Bayesian sense, yet the final verdict is \emph{never} dependent on any \emph{ad hoc} suppositions. This immediately motivates the goal of searching for the truncation basis that maximizes the relative-belief ratio in the quest for finding the truly smallest Hilbert space containing the unknown state.

\begin{acknowledgments}
The authors thank J.~Sperling and J.~Gil-Lopez for insightful discussions. This work was supported by the European Union’s Horizon 2020 Research and Innovation Programme Grant No. 899587 (Project Stormytune). Y.S.T., S.U.S. and H.J. acknowledge support from the National Research Foundation of Korea (NRF) grants funded by the Korean government (Grant Nos. NRF2020R1A2C1008609, 2023R1A2C1006115, NRF2022M3E4A1076099 and RS-2023-00237959) \emph{via} the Institute of Applied Physics at Seoul National University, the Institute of Information \& Communications Technology Planning \& Evaluation (IITP) grant funded by the Korea government (MSIT) (IITP-2021-0-01059 and IITP-20232020-0-01606), and the Brain Korea 21 FOUR Project grant funded by the Korean Ministry of Education. M.E. was supported by a grant form the Natural Sciences and Engineering Research Council of Canada 2017-06758. L.L.S.S.  acknowledges support from Ministerio de Ciencia e Innovaci\'on (Grant PID2021-127781NB-I00).
\end{acknowledgments}

\clearpage
\onecolumngrid
\begin{center}
	\large \bf Supplemental Material
\end{center}
\appendix

\section{Maximum-likelihood estimators constrained with background-noise subtraction}

The ML state estimators~$\ML(d)$ are found using a highly efficient numerical procedure that maximizes the log-likelihood using projected-gradient methods~\cite{Shang:2017sf}. Instead of walking the $A$~space through the parametrization $\rho=A^\dag A/\tr{A^\dag A}$, which results in search trajectories that hovers (zig-zags) around the global maximum, projected-gradient recipes suggest an iteration of two steps: an optimization update in the $\rho$-space followed by a projection of the unphysical result back to the state space. This projection is done by setting all negative eigenvalues to zero and a final trace renormalization.

\begin{figure}[h!]
	\includegraphics[width=0.5\columnwidth]{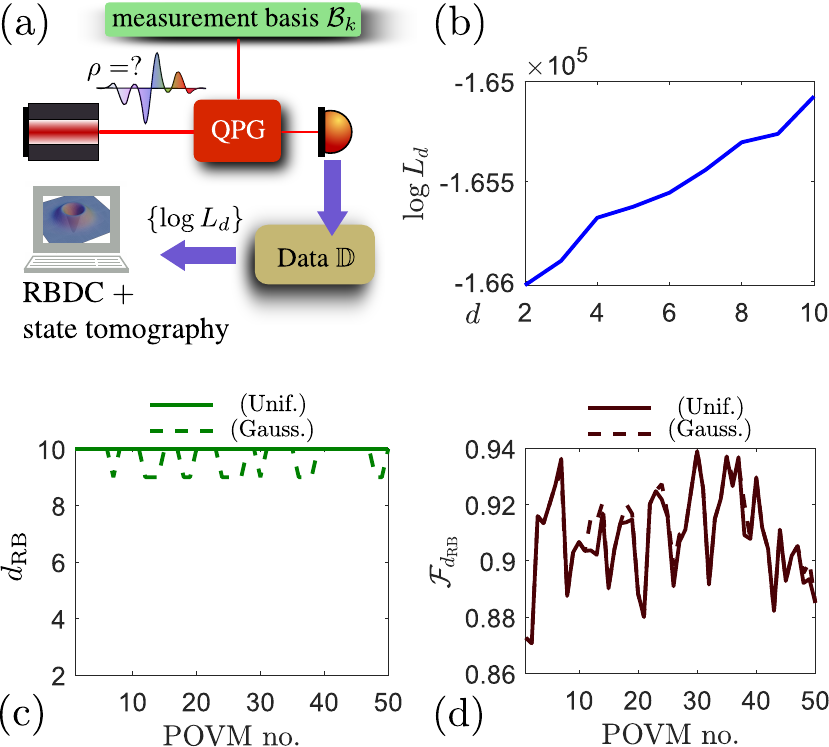}
	\caption{\label{fig:TF_bkgd}(a)~The experimental scheme for RBDC tomography on the HG$_1$ temporal quantum state using the same 50~POVMs in Fig.~2 of the main text. All specifications are the same except that no background noise is subtracted. (b)~All RB~ratios are below~1 in this case. Thus, all dimensions are ruled out by the data due to lack of evidence. (c)~This time, the $\dRB$s are larger without background-noise subtraction. (d)~The fidelities of the reconstructed states are, of course, generally lower now.}
\end{figure}

Additional simple constraints may be flexibly incorporated to augment the current projected-gradient procedure. In our context, we are interested in finding ML estimators that are also void of small diagonal-element biases (smaller than a particular threshold of, say, 0.01). After the state-space projection, we remove such biases by setting all diagonal elements of $\ML(d)$ that are smaller than this threshold value and their corresponding row and column elements to zero, followed by a trace renormalization. It is straightforward to see that positivity is still preserved, and so background-noise subtraction performed this way is therefore compatible with the above state-space projection. The resulting $\ML(d)$ shall therefore be physical and possess no diagonal-element bias.

We note that the imposition of this bias-free constraint would result in an ML problem that is nonconvex. In order to find the global maximum when $d=2$ (lowest dimension), we repeat the search multiple times to obtain a log-likelihood value that is heuristically close to the global maximum one. We do the same for $d>2$, but ensure that the maximal value should at least be equal (up to numerical precision, of course) to the previous lower dimension. If all ML estimators gave lower log-likelihood values, then we take the estimator for the previous dimension, pad it with zeros in the computational basis and consider this as the ML estimator for this dimension.

\section{Data with background noise (systematic errors)}

It is clear that RBDC pays no attention to the type of datasets obtained in any experiment. It treats all datasets on the same footing and analyze them using the key elements found in Bayesian statistics: So long as the posterior distribution is larger than the prior for a particular dimension~$d$, then the particular dataset is deemed to support the statement that a $d$-dimensional Hilbert space contains the unknown state~$\rho$.

Figure~\ref{fig:TF_bkgd} shows that if one applies RBDC on datasets containing background noise, then the effective dimension~$\dRB$ predicted by RBDC will generally be large. With a (non-pathological) prior that reflects a stronger belief for a particular dimension, and if that dimension coincides with the true support dimension, then $\dRB$ can be slightly smaller, partially subject to the verdict of the dataset according to the posterior probability. Even in this case, RBDC always consults the dataset, and only the dataset. No other spurious assumptions are made in such a dimension certification.

That $\dRB$ is consistently large in this situation is \emph{not} a drawback. Rather, it is a warning to the observer that the dataset holds systematic errors---overly optimistic model-selection procedures can lead to unjustified assertions than what the dataset dictates.

\section{Table of numerical values}

Tables of likelihood values and posterior contents used to plot Figs.~2(b) and 3(b) in the main text. The decimal places showcase storage capabilities, \textbf{not} computational precision which is dictated by that of ML.

\begin{table}[h!]
	
	\caption{\label{tab:fig2main}Likelihood values and posterior probabilities ffrom the dataset of POVM~no.~7 for Fig.~2(b) in the text.}
	\vspace{2ex}
	\begin{tabular}{ll}
		\textbf{Likelihood} &
		\begin{tabular}{lr}
			\hline
			$d$ & $L_d$ \\[2ex]
			& \hphantom{ 0.000000000001877816258434481807624919489531665960511015092814392901090475765}\\[-4ex]
			2 & 1.910200961849743785030610224731732032006327450641135205959787153e-149752\\
			3 & 5.106857911370325091648503512521967863132578983385229925814312374e-149658\\
			4 & 1.794717885295998128537139948794055184939542610517891728092474606e-149659\\
			5 & 6.755446025787587091812436483073267911850465319370140060645363117e-149625\\
			6 & 6.755446025787587091812436483073267911850465319370140060645363117e-149625\\
			7 & 6.755446025787587091812436483073267911850465319370140060645363117e-149625\\
			8 & 6.755446025787587091812436483073267911850465319370140060645363117e-149625\\
			9 & 6.755446025787587091812436483073267911850465319370140060645363117e-149625\\
			10 & 6.755446025787587091812436483073267911850465319370140060645363117e-149625
		\end{tabular}\\	
		\textbf{Uniform} & 
		\begin{tabular}{lr}
			\hline
			$d$ & $\PR(d|\mathbb{D})$ \\[2ex]
			& \hphantom{ 0.000000000001877816258434481807624919489531665960511015092814392901090475765}\\[-4ex]
			2 & 4.712743255732553054521532547018926651753778872759252784183438783e-129\\
			3 & 1.259936030839883114653205748522088712368481643488005315058207893e-34\\
			4 & 4.427829730376093205762548271647074566243289026592105496299827386e-36\\
			5 & 0.1666666666666666666666666666666666449297611976059325548194794794\\
			6 & 0.1666666666666666666666666666666666449297611976059325548194794794\\
			7 & 0.1666666666666666666666666666666666449297611976059325548194794794\\
			8 & 0.1666666666666666666666666666666666449297611976059325548194794794\\
			9 & 0.1666666666666666666666666666666666449297611976059325548194794794\\
			10 & 0.1666666666666666666666666666666666449297611976059325548194794794
		\end{tabular}\\	
		\textbf{Gaussian} &
		\begin{tabular}{lr}
			\hline
			$d$ & $\PR(d|\mathbb{D})$ \\[2ex]
			2 & 1.811492525702610273619644632523871842090688357968370020285770368e-123\\
			3 & 1.781626975790019495910583089918198454779828033023700911111053681e-29\\
			4 & 2.917819206131510044602153241393249605968452591884373208503656677e-32\\
			5 & 0.9990888364735183604306543028149505613243627440237335371178507634\\
			6 & 0.0009110510919670464577397712746462635210025962523776749399412791270\\
			7 & 0.0000001124326367726086868358410450581788270436362962270528277584970265\\
			8 & 0.000000000001877816258434481807624919489531665960511015092814392901090475765\\
			9 & 4.244483309846593936193234147159523777178454647022659632446218542e-18\\
			10 & 1.298397293813343080292908005916186515242034448853146959049695543e-24\\
			\hline
		\end{tabular}
	\end{tabular}
\end{table}

\begin{table}[h!]
	
	\caption{\label{tab:fig3Bellmain}Likelihood values and posterior contents from different priors for the Bell-state combined dataset of all nine polarization settings used in Fig.~3(b) of the main text.}
	\vspace{2ex}
	\begin{tabular}{ll}
		\textbf{Likelihood} &
		\begin{tabular}{lr}
			\hline
			$d$ & $L_d$ \\[2ex]
			& \hphantom{}\\[-4ex]
			2 & 8.403490385863592712990917327109636785426579254762189116001695376e-241771739\\
			3 & 3.655739282866103174044744995755057155229460506918790692749954769e-198298861\\
			4 & 6.330620382524476308070544809567671044319190406322579287081083635e-196822260\\
			5 & 1.776792027639825604150877605451397800972526507582163108814115906e-196778830\\
			6 & 1.776792027639825604150877605451397800972526507582163108814115906e-196778830\\
			7 & 1.776792027639825604150877605451397800972526507582163108814115906e-196778830\\
			8 & 1.776792027639825604150877605451397800972526507582163108814115906e-196778830\\
			9 & 1.776792027639825604150877605451397800972526507582163108814115906e-196778830\\
		\end{tabular}\\	
		\textbf{Uniform} & 
		\begin{tabular}{lr}
			\hline
			$d$ & $\PR(d|\mathbb{D})$ \\[2ex]
			& \hphantom{8.403490385863592712990917327109636785426579254762189116001695376e-241771739}\\[-4ex]
			2 & 9.459171647709653534886755317970275165136876060380205043839921333e-44992910\\
			3 & 4.114988390309414313454661644948699303165029949647399242731604008e-1520032\\
			4 & 7.125899130618746242142348837497304737217532444365261039751559732e-43431\\
			5 & 0.2\\
			6 & 0.2\\
			7 & 0.2\\
			8 & 0.2\\
			9 & 0.2\\
		\end{tabular}\\	
		\textbf{Gaussian} &
		\begin{tabular}{lr}
			\hline
			$d$ & $\PR(d|\mathbb{D})$ \\[2ex]
			& \hphantom{8.403490385863592712990917327109636785426579254762189116001695376e-241771739}\\[-4ex]
			2 & 4.210267819553408908553507147335813649770725713580218068865734232e-44992913\\
			3 & 2.718301595960439593842719321802808035882597596336721506037869889e-1520033\\
			4 & 9.454795548003367249196222449106976134248370921079357367617165407e-43431\\
			5 & 0.7213349069032195596568817076250338665875643824181071226498860823\\
			6 & 0.2653642824490107955366354177322951490485674592347429120982628122\\
			7 & 0.01321170967267805579371765694826247023434539773236961482957872417\\
			8 & 0.00008901979954180955844229495404517222736201531945568018301054324833\\
			9 & 0.00000008117554977945432292274036334190216074529532467023926183802934895\\
			\hline
		\end{tabular}
	\end{tabular}
\end{table}

\begin{table}[h!]
	
	\caption{\label{tab:fig3TMSVmain}Likelihood values and posterior contents from different priors for the TMSV-state combined dataset of all nine polarization settings used in Fig.~3(b) of the main text.}
	\vspace{2ex}
	\begin{tabular}{ll}
		\textbf{Likelihood} &
		\begin{tabular}{lr}
			\hline
			$d$ & $L_d$ \\[2ex]
			& \hphantom{}\\[-4ex]
			2 & 1.222256227141671752091154406796389699070660098072229744455202464e-306872481\\
			3 & 2.844096023090229079813918119779081530244493176898802665681025320e-230480082\\
			4 & 8.480194198000157649647279087686896087251218208326448855277658249e-226267426\\
			5 & 1.476921537445382768539500507106717819644837744873381994693482444e-226050278\\
			6 & 4.145221565372174318775201702865502600615951978650069696694988681e-226006849\\
			7 & 4.145221565372174318775201702865502600615951978650069696694988681e-226006849\\
			8 & 4.145221565372174318775201702865502600615951978650069696694988681e-226006849\\
			9 & 4.145221565372174318775201702865502600615951978650069696694988681e-226006849\\
		\end{tabular}\\	
		\textbf{Uniform} & 
		\begin{tabular}{lr}
			\hline
			$d$ & $\PR(d|\mathbb{D})$ \\[2ex]
			& \hphantom{1.222256227141671752091154406796389699070660098072229744455202464e-306872481}\\[-4ex]
			2 & 7.371477060189017748930867093088421250793582773000497335680827786e-80865634\\
			3 & 1.715285888967237229526402291882300354038778693576490201836411362e-4473234\\
			4 & 5.114439641080302793122772178037966165638469533725511505894716085e-260578\\
			5 & 8.907373913273432802677936046871630921521915555456576299689449665e-43431\\
			6 & 0.25\\
			7 & 0.25\\
			8 & 0.25\\
			9 & 0.25\\
		\end{tabular}\\	
		\textbf{Gaussian} &
		\begin{tabular}{lr}
			\hline
			$d$ & $\PR(d|\mathbb{D})$ \\[2ex]
			& \hphantom{1.222256227141671752091154406796389699070660098072229744455202464e-306872481}\\[-4ex]
			2 & 9.419298311571059005750921671935084393211215810269869590987778579e-80865637\\
			3 & 3.252916814063418799244303185375878854919880381240512563418771974e-4473235\\
			4 & 1.948130051599535872125222000144826411876368089914200843629733303e-260577\\
			5 & 9.222826814913319689845696149798283514283166867987880881502717475e-43430\\
			6 & 0.9522695487261826844081002875207288867423403130685955483059846066\\
			7 & 0.04741070912706539274878167382556835549396161713625131659096852837\\
			8 & 0.0003194508452872242773780641178805601027960203778077992646929174449\\
			9 & 0.0000002913014646985657399745358221976609020494173453358383539476069818\\
			\hline
		\end{tabular}
	\end{tabular}
\end{table}

\end{document}